# Poisoning the Evidence: Document and Data Attacks on LLM-driven Retrieval-Assisted Analytical Process-Property Model Generation

Hoang Anh Nguyen[1], Ziqin Liu[2], Yuan Hong[2], Hongyi Xu[1*]
[1]School of Mechanical, Aerospace, and Manufacturing Engineering, University of Connecticut, Storrs, CT 06269
[2]School of Computing, University of Connecticut, Storrs, CT 06269
* Corresponding Author: hongyi.3.xu@uconn.edu

## Abstract

Large Language Model (LLM)-driven, retrieval-assisted workflows generate closed-form process-property analytical models by integrating technical literature with small experimental datasets. However, internal agreement between the two sources, literature and dataset, can create a false sense of model validity if both inputs are compromised. We propose a new coordinated dual-channel poisoning approach that constructs a poisoned document and bounded alterations to a numerical dataset from the same model coefficient alteration, such that the altered analytical model achieves the same level of statistical agreement with the poisoned dataset as the baseline model performed with the unpoisoned dataset. We evaluate attack propagation through retrieval, generation, fitting, and selection, accessing prediction accuracy against original clean benchmark data kept outside the attacker's access. Evaluated on a flow assisted laser-induced plasma micro-machining process benchmark, the attack successfully forces poisoned literature claims into retrieved summaries and shifts final process predictions at chosen operating conditions, significantly degrading clean-test accuracy. Furthermore, because coefficient refitting preserves baseline residuals, standard internal diagnostic checks fail to alert users to the compromise. These findings demonstrate that internal consistency between documentation and empirical data cannot substitute for independent model verification against protected ground truth.

## 1. Introduction

Digital manufacturing connects digital design, process control, and quality inspection into interconnected data streams. While this connectivity enables seamless knowledge sharing and automated decision-making across the production pipeline, it create paths for cyber-physical attacks, which corrupted digital information to propagate into physical manufacturing workflow [1-4].

Large language models (LLMs) are increasingly being investigated for manufacturing knowledge support and decision-making [5]. Recent applications include query-adaptive retrieval of manufacturing knowledge [6], manufacturing task planning [7], engineering document interpretation [8] and interpretation of mechanical experiments and additive-manufacturing quality [9-11]. As Large Language Models (LLMs)-driven, retrieval-assisted Process-Property predictve model generation becomes an emerging paradigm for constructing process-property relationships from engineering literature and small experimental datasets [12, 13], these cyber-physical vulnerabilities extend into the Artificial Intelligence (AI) generation pipeline itself.

State-of-the-art LLM-driven Process-Property modeling leverages Retrieval-Augmented Generation (RAG) [14] to integrate physical domain knowledge with experimental measurement data. These LLM-driven frameworks retrieve governing principles from engineering literature to propose candidate closed-form equations, while experimental data fit their numerical coefficients and guide the iterative refinement of the analytical function forms [13, 15]. In this study, we use this state-of-the-art RAG-LLM modeling framework as our benchmark system to investigate its vulnerabilities to coordinated dual-channel poisoning attacks.

Integrating retrieved text into the model generation pipeline expands the attack surface [16, 17]. Benchmarks such as InjecAgent and AgentDojo show that malicious external content can redirect tool-integrated LLM agents [18, 19]. In addition to the well-established thread of corrupting training data [20], an adversary can manipulate the predicted input-output relationship through altering the engineering documents for information retrieval. When a compromised document and corrupted data agree on the same flawed relationship, their alignment creates a false sense of corroboration despite their shared adversarial origin. Therefore, this thread stealthily compromises the prediction model during its generation phase, before it is deployed to guide downstream decisions.

Existing research addresses threats and detection at several stages in this information pipeline. In manufacturing security, previous studies examine corrupted production files, attack detection, and process resilience [1, 2, 4, 21, 22]. In AI security, prior work examines poisoned retrieval text and code [17, 23, 24], regression poisoning, and false-data injection [20, 25, 26]. Albeit these results motivate the present study, a critical research gap still remains: they do not address how poisoned claims (e.g., the parametric relationships derived from literature) propagate through the full model-generation pipeline, from equation proposal and numerical fitting to final model selection. In such multi-stage architectures, an injected claim might be filtered out or numerically altered during the iterative model refinement process.

Manufacturing applications therefore require a systematic way to assess whether a AI-generated process model is supported by trustworthy evidence. This assessment is especially critical when models extrapolate to operating conditions outside their calibration range, where the AI-generated predictive models extend beyond the available observations. This study investigates how coordinated corruption of documents and training data alters the selected model's predictions, and what evidence is required to detect such changes. This work makes three new contributions:

- Coordinated poisoning attack construction: We formulate a poisoning framework where a document-based process-property relationship claim and bounded alterations to experiment dataset (for training) jointly represent the same poisoned analytical model, without modifying the underlying model generation procedure.
- Pipeline-wide evaluation approach: We track the propagation of the adversarial alterations through retrieval, synthesis, fitting, and selection. By comparing clean, document-only, data-only, and coordinated poisoning conditions, we can distinguish whether the candidate analytical Process-Property model selected by LLM simply because the text repeats it, or because the dataset actually supports it.
- False sense of model validity: We showed that a poisoned model can easily pass internal diagnostic evaluation while still suffering high error on original clean test data. Because the altered text and poisoned experimental data agree with each other, standard diagnostic checks detect no issue. We demonstrate the importance of using uncorrupted, held-out ground-truth data to reveal the model's true predictive errors.

The remainder of this paper is organized as follows. Section 2 reviews the background and related work. Section 3 presents the coordinated document and data poisoning method, and Section 4 describes the experimental setup. Section 5 reports the poisoning effects and diagnostic findings. Section 6 summarizes the main findings and discusses limitations and future work.

## 2. Background and Related Work

Figure 1 highlights the role of predictive models within three connected layers: physical manufacturing, manufacturing information, and digital modeling and decision support. Manufacturing-security research establishes that corrupted engineering information can affect production. Sturm et al. exam ine attacks on additive-manufacturing files [1], and Belikovetsky et al. demonstrate a malicious design change that reduces a printed propeller's fatigue life and causes failure [3]. Physics based monitoring offers another perspective by checking machining behavior during production [2]. The background reviewed below explains how engineering documents and calibration data propagates into the generation of the predictive model and highlights integrity threats studied in prior works.

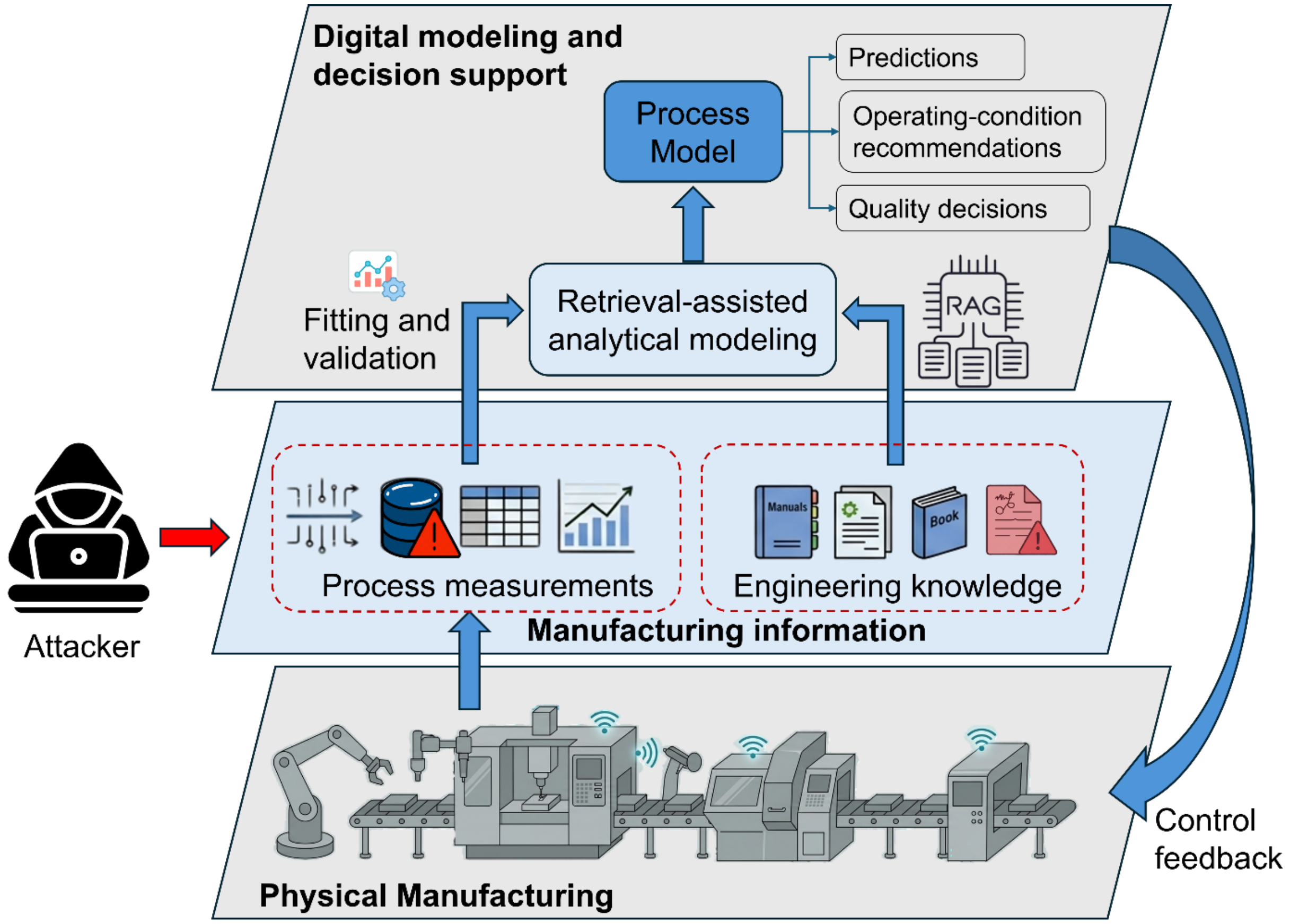

**Figure 1.** Manufacturing information links physical production to digital modeling and decision support. Red dashed boxes mark the two input channels: process measurements and engineering knowledge. RAG denotes retrieval-augmented generation. The feedback loop illustrates potential use; physical effects and closed-loop control are outside this evaluation.

*2.1 Large language models and retrieval-augmented generation*

Large language models (LLMs) can generate task-specific responses via prompt instructions and few-shot examples without requiring model retraining [27]. A recent review examines how language models extract relationships among processing, material structure, and properties from technical literature [28]. In manufacturing applications, effective responses heavily depend on domain-specific context, such as engineering reports and experimental data. Retrieval-augmented generation (RAG) fulfill this need by integrating relevant external documents directly into the prompt context during generation [14].

In a standard RAG pipeline, a document collection is split into passages and indexed for efficient retrieval. Given a query, the system retrieves relevant passages, often refined by a reranker before feeding the passages into the LLM's prompt alongside the query question. For analytical model generation, the LLM uses this retrieved context to synthesize candidate functional relationships and process trends [29].

This LLM synthesis step connects source documents to downstream model generation decisions. When a proposed relationship is retained in the synthesized summary, it becomes available to the equation generation stage. However, while retrieval evaluates relevance to the query, it does not guarantee the factual accuracy of a source, leaving the pipeline vulnerable to knowledge-poisoning attacks [17]. Numerical experimental data subsequently play a separate role: calibrating the analytical model coefficient and evaluate the accuracy of the generated analytical models [29].

*2.2 Retrieval-assisted analytical model generation*

An analytical process model expresses the engineering response as an explicit function of process parameters. Among the simplest classical approaches, Response Surface Methods require the analyst to select a polynomial function form and fits its coefficients on observation data [30]. To automate the function form discovery, Symbolic Regression searches for the explicit function form directly. An latest development in SR is LLM-SR, which uses an LLM's scientific prior knowledge to propose equation programs, combines these proposals with data-driven parameter optimization, and refines the search using previously evaluated candidates [29]. In manufacturing, LLM-based agents also coordinate data analysis and model selection for prescriptive maintenance [31].

Retrieval introduces an external source of knowledge to model generation. Naghavi Khanghah et al. combine automated knowledge extraction from manufacturing literature with iterative equation refinement [15]. The knowledge summary describes relationships that can guide the generation of candidate analytical Process-Property models. Coefficients of the analytical models are obtained by fitting on a small experimental dataset, while predictive performance metrics (e.g., R2) provide feedback to refine subsequent equation generations.

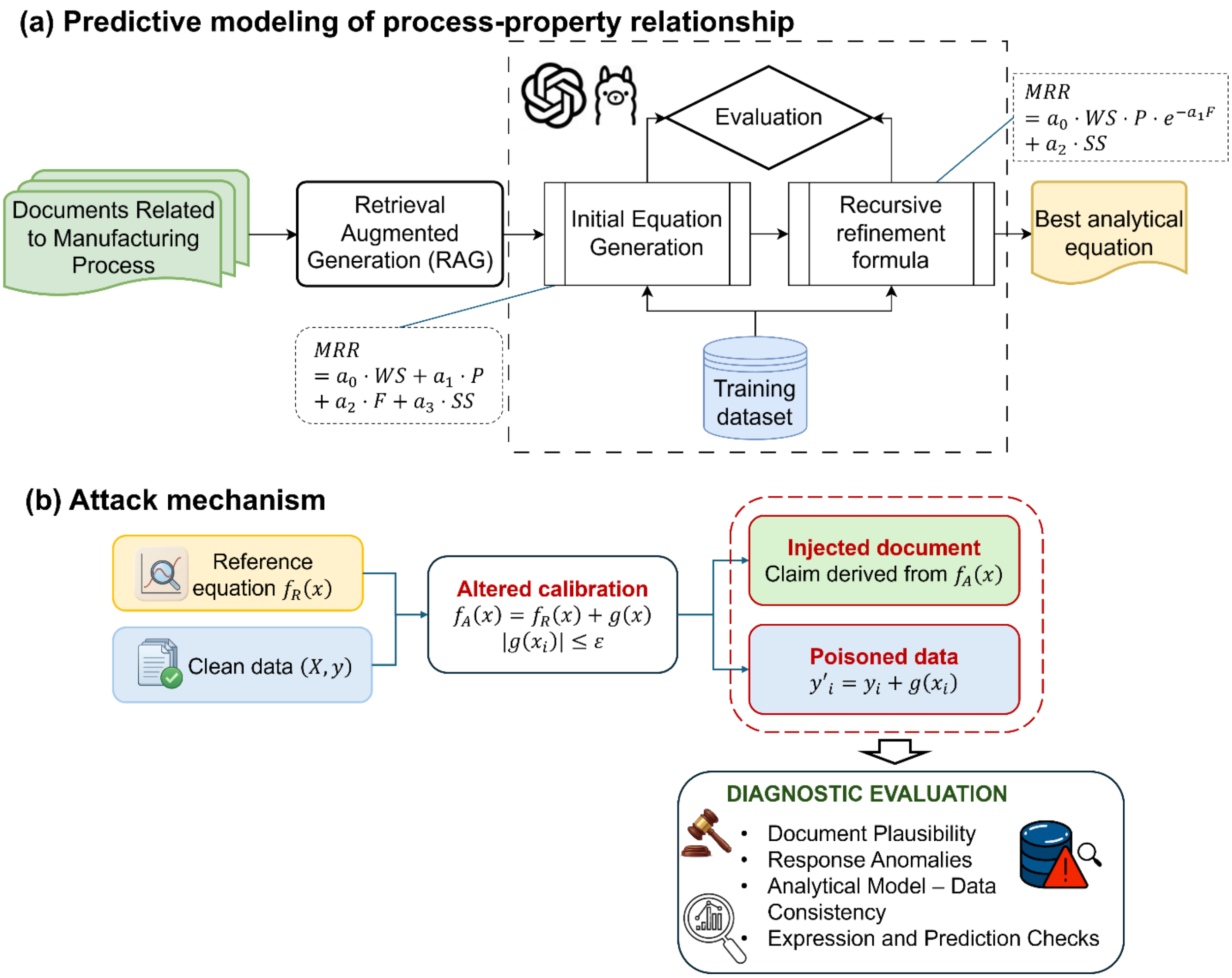

**Figure 2.** Retrieval-assisted predictive modeling (a) and coordinated poisoning (b). An altered analytical model defines the poisoned source document and poisoned data. Diagnostic checks assess the inputs and selected analytical models without filtering the search. Model icons and example equations are schematic.

In this architecture, the LLM's proposed parametric relationships and the dataset jointly determine the selection of candidate analytical models. Naghavi Khanghah et al. evaluate this interaction through interpolation and extrapolation accuracies performance across manufacturing processes and comparisons of knowledge extraction and refinement [15].

### *2.3 Existing attacks on retrieved knowledge and calibration data*

Retrieval-augmented language models are vulnerable to upstream text manipulation via both prompt instructions and factual claim. Greshake et al. demonstrate indirect prompt injection through external material ingested during retrieval [16]. PoisonedRAG shows that strategically inserted documents can force an LLM to generate specific attacker-chosen answer [17]. AuthChain further studies attacks through a single poisoned document [23], and Lin et al. examine how poisoned retrieved code examples introduce vulnerabilities into generated code [24]. While these studies evaluate attack success based on generated natural language text, they do not examine how a corrupted text claim propagates when passed into a downstream analytical model generator that relies on parameter fitting and model selection.

On the other hand, data-poisoning studies focus on corrupting the observation data (e.g. experimental dataset), which is used to calibrate the generated analytical models. Jagielski et al. study attacks on linear regression and protection through robust estimation [20]. Carnerero-Cano et al. examine the effectiveness-detectability trade-off for linear and nonlinear regressors and introduce BayesClean [32]. While these studies quantify how corrupted measurements degrade predictive models, they assume fixed mathematical model structures and omit textual retrieval channels altogether.

There also exist prior works that explain why internal diagnostic consistency checks can fail to detect corruption. Liu et al. construct false-data-injection attacks that preserve state-estimation residuals, the signed differences between the ground-truth values and fitted model predictions [25], and Russo studies poisoning and detectability in least-squares identification [26]. While residual preservation is an established thread in control and state estimation, its implications for retrieval-assisted process modeling remain unexplored. Specifically, prior work leaves open how a coordinated alteration that simultaneously injected into literature claims and experimental data propagates through candidate analytical model generation and selection, also how these alterations affect prediction accuracy when predictions are compared with ground-truth values from an original clean dataset.

## 3. Attack mechanism

This section formalizes the proposed attack mechanism, detailing the adversary's access requirements, the coordinated dual-channel poisoning attack depicted in Figure 2(b), and the evidence available to diagnostic checks of the two input channels. The attack begins with a correct model generated from the baseline LLM-driven process-property model generation framework and applies a single, chosen mathematical perturbation to simultaneously alter both the retrieved literature claims and the experimental dataset in a synchronized way.

### 3.1 Threat model

Following standard regression-poisoning frameworks, we define the threat model through the attacker's objective, knowledge of the modeling workflow, and capability to alter inputs [20].

The attacker's objective is to induce a maximized change in the predicted output at a specified combination of input process-variable values $x_t$, subject to the permitted input alterations. This objective concerns the final analytical model returned after the workflow generates candidate analytical models, fits their coefficients, and selects a candidate. The prediction change is assessed relative to the same workflow using original unpoisoned documents and an original clean experimental dataset.

To prepare the attack, the attacker observes the process query and recorded process-variable values, chooses the process-variable values$x_t$, and inspects an original clean reference analytical model and its coefficients. The attack does not require access to the true physical governing law or advance knowledge of which candidate analytical model the workflow will ultimately select.

The attacker has the capability to access and alter both the source documents and experimental dataset before they are imported into the workflow. Consider a shared workspace containing calibration spreadsheets, source documents, and saved analytical models. An insider or compromised account with permission to alter these inputs could poison them. Document access allows the attacker to add or alter a source document imported into the retrieval collection, an insider setting also considered by PoisonedRAG [17]. Dataset access allows alterations to recorded output values without requiring control of measurement equipment. Coordinated dual-channel poisoning requires write access to both sources; each single-channel attack requires access only to its source. We assume that the entire dataset imported into the workflow can be altered, including its training, validation, and test datasets. Process-variable values remain fixed, and every alteration to a recorded output value is bounded. A clean copy of the dataset is saved outside the attacker's access as the ground-truth reference for assessing prediction accuracy. The attacker also do not have the access to the original modeling workflow, including retrieval settings, prompt instructions, language-model configuration, fitting and selection rules, and diagnostic implementations. The mechanisms used to compromise these permissions fall under general network and enterprise IT security and are outside this study's scope.

### *3.2 Modeling workflow and attack overview*

We evaluate the attack in a retrieval-assisted predictive modeling workflow adapted from Naghavi Khanghah et al. [15], shown in Figure 2(a). The initial generation asks the LLM to propose analytical models with adjustable coefficients from the knowledge summary. Each candidate is fitted to training dataset by numerical least squares and scored on a validation subset. Ranking combines normalized validation mean squared error (MSE) with a penalty for expression size and parameter count. Candidates that cannot be evaluated or fitted are excluded. The calibration table enters through numerical fitting and evaluation; its raw outputs are not supplied directly in the initial equation-proposal prompt.

Refinement returns leading models, their validation MSE, and their coefficient of determination ($R^2$) to the LLM alongside the knowledge summary. New proposals undergo the same fitting and ranking process until a stopping criterion or search budget is reached. The final output is the best-ranked surviving model with fitted coefficients. Held-out test data and the attacker's chosen process-variable data points are evaluated after selection.

Coordination alteration is necessary because a document claim and the experimental data govern different stages of the model generation pipeline. While a retrieved claim guides the structural formulation of candidate analytical models, subsequent numerical parameter fitting against experimental data can easily reject an ill-fitting structure in favor of a different candidate. To prevent such diagnostic rejection, the coordinated attack aligns these two sources by presenting an altered model in the document and applying its prediction changes to the experimental data table.

Because the system is free to discard the attacker's suggested form during iterative generation, attack success is evaluated based on the final model generated by the framework.

*3.3 Designing the altered analytical model*

The attacker initiates the construction by selecting a saved reference process model, denoted $f_R$, which defines the analytical function terms whose coefficients can be altered. Because $f_R$ is an empirical model rather than an exact physical law, fitting it to the original clean training dataset does not guarantee that it describes the physical process exactly.

The mathematical coupling between coefficient alterations and training data alterations follows the residual-preserving principle of false-data injection. Liu et al. show that targeted shifts to sensor measurements can be accompanied by data alterations that leave the linear estimator's residual invariant [25]. In our context, the fixed analytical model functions map coefficient alterations to prediction changes. Writing this prediction change as $g(x)$ at process setting $x$, the altered analytical model becomes: $f_A(x) = f_R(x) + g(x)$. If an output value $y_i$ is changed to $y_i' = y_i + g(x_i)$, its associate residual, defined as dataset output minus prediction, satisfies $y_i' - f_A(x_i) = y_i - f_R(x_i)$.

To operate the attack, the attacker chooses a maximum alteration allowance of edit bound $\epsilon \geq 0$ for each dataset output value, defined in the same physical units as the output in the experimental dataset, and so $| y_i' - y_i | = | g(x_i) | \leq \epsilon$ for every dataset entry *i*. Constraining alteration magnitude draws precedent from physical system identification, where Russo applies relative Frobenius-norm bounds to experimental dataset modifications [25]. Our adaptation applies the bound to each stored output entry.

Within this bound, the attacker optimizes coefficient alterations that maximizes the change in the final altered model's prediction at $x_t$ relative to the original clean reference analytical model. Meyer et al. analyze worst-case prediction changes at a fixed input under individually bounded alterations to training output values [33]. Our construction additionally requires these alterations to follow a shared analytical model coefficient vector, which also determines the poisoned source document. Let $\phi(x)$ collect the fixed model function terms and let $\Delta\beta$ denote their real-valued coefficient alterations, so $g(x) = \phi(x)^T \Delta\beta$. For a dataset with N entries, the attacker solves

$$\begin{aligned} & | g(x_t) | = | f_A(x_t) - f_R(x_t) | = | \phi(x_t)^T \Delta\beta | \\ \text{subject to } & | g(x_i) | = | \phi(x_i)^T \Delta\beta | \leq \epsilon, \; i = 1, \ldots, N. \end{aligned} \tag{1}$$

The objective maximizes the magnitude of the model prediction change at the inputs $x_t$ chosen by the attacker, $x_t \neq x_i$. The constraints instead limit the alterations to dataset output values at the inputs $x_i$, since $g(x_i) = y_i' - y_i$. In the interpolation case where $x_t$ lies inside the ranges covered by the dataset, each output alteration is allowed to be anywhere between $-\epsilon$ and $+\epsilon$, including the endpoints. For manufacturing decision, even a small model output changes during interpolation can affect which process-variable values are preferred. Combinations with similar predicted performance may change rank, or a prediction may cross a specified performance threshold. In extrapolation beyond the dataset range, the prediction change can exceed $\epsilon$, and the alteration can make the poisoned model underpredicting the final output.

Notably, this formulation remains linear with respect to the coefficient alterations $\Delta\beta$ even when the terms $\phi(x)$ contain non-linear functions of the process variables. The resulting optimal vector $\Delta\beta^*$ defines the corrupted model $f_A(x)$ used to generate both the textual document claim and the modified experimental dataset. Section 4.2 demonstrates the solution for the benchmark problem.

*3.4 Document poisoning*

The document channel injects the poisoned analytical $f_A$ into the RAG literature knowledge retrieval used to guide downstream analytical model proposal. Adapting the knowledge-corruption mechanism from PoisonedRAG, the attack relies on inserting an engineered passage that is both successfully retrieved and incorporated into the synthesized knowledge summary [17]. Unlike PoisonedRAG's iterative text optimization, our approach directly constructs a poisoned source document describing $f_A$ and its predictions, from the same coefficient alteration $\Delta\beta$ used for the bounded edits to experimental data.

The core deception involves false provenance attribution. The source document begins with the statement: "This document reports the reference analytical fitted to the unmodified training data before the current analysis." It then substitutes $f_A$ for the genuine reference analytical model $f_R$, making the poisoned model appear to be a validated historical model from literature. The associated poisoned explanation text defines the process variables and units, describes their effects on process output value, and reports the prediction at $x_t$.

To evaluate how structural context affects retrieval and synthesis, we construct three source document variants based on the model text I and process query Q. These variants adapt PoisonedRAG's content-only baseline, question-prefixed attack, and reversed-order ablation [17]:

- Document Alone (I): Contains only the poisoned source document text.
- Question-Document (Q+I): Prepends the query Q before I.
- Document-Question (I + Q): Appends Q after I as an ordering ablation.

All three variants contain the same poisoned analytical model $f_A$, explanation of process relationships, and predicted output at the chosen process-variable values $x_t$. Each begins with the same heading marking the text as a synthetic benchmark document. It is noted that the heading only marks the added fake documents for human validation of the attack model and is not used for detection. In each document, only the presence and position of the process query differ, allowing us to examine their effects on retrieval and synthesis.

*3.5. Data poisoning through bounded dataset alterations*

The data channel alters the training dataset used to fit and rank the proposed analytical models. Before the data table is imported, the attacker adds $g(x_i)$ to the original output value $y_i$ at each setting, yielding the modified output $y_i' = y_i + g(x_i)$. The process settings remain unchanged.

Each edited output value $y_i$ retains its original residual relative to the poisoned model $f_A(x)$. For example, adding $g(x_i)$ to a value 1 μm³/s above $f_R(x_i)$ places it 1 μm³/s above $f_A(x_i)$.

*3.6 Attack detectability and diagnose metrics*

Attack detectability depends on whether a defense diagnostic evaluates only the supplied inputs or has access to protected ground-truth reference data. False-data-injection and system-identification studies show that altered data can remain compatible with a fitted model and evade residual-based tests [25, 26]. A parallel limitation exists in natural language processing: frameworks such as RAGAS assess LLM faithfulness by verifying whether claims align with the retrieved context [34]. Under our threat model, because the text corpus itself is compromised, validating a candidate model against local documentation cannot establish its true physical authenticity. Table 1 summarizes four diagnostic categories evaluated in this work.

Table 1. Information used and purpose of the four diagnostics.

| Diagnostic | Information used | What is assessed |
| --- | --- | --- |
| D1: Document Plausibility [35] | Poisoned model source document, original corpus, and process description; no protected reference analytical model. | Consistency and plausibility of the document, rated by an LLM across document orders. |
| D2: Response Anomalies [20, 36-38] | Process settings and training/validation dataset; original-table rankings provide the evaluation control. | Which rows the statistical outlier diagnostics prioritize and how data sample alterations change that ranking. |
| D3: Analytical Model-Data Consistency [25, 26] | Candidate equations and coefficients, fixed reference terms, process settings, and experimental data. | Analytical model-data residuals before refitting and after unconstrained ordinary least squares using the same terms. |

| Diagnostic | Information used | What is assessed |
|---|---|---|
| D4: Expression and Prediction Checks [39] | Selected analytical models and evaluation-grid settings; original-output magnitudes set the limit for extreme predictions. | Expression structure, nonfinite or extreme predictions, and excessive constant magnitudes, or extreme extrapolation values. |

D1 Document Plausibility: D1 evaluates the credibility of the source document (i.e. malicious document describing $f_A$ and its predictions) used to guide equation proposals. Zheng et al. study LLM scoring of assistant answers and report position bias in judging [35]. We adapt this scoring approach to assess consistency with the poisoned corpus and process description, without access a reference equation. We summarize the scores and inspect the judge's reasons to examine how plausible the corrupted text appears. These scores represent heuristic suspicion rating rather than calibrated probability of poisoning metrics or formal rejection thresholds.

D2 Response Anomalies: D2 tests whether point-wise output modifications can be detected by statistical outlier diagnostics. We evaluate standard regression quality metrics and outlier toolkits including cleanlab regression label-quality scores [36], Cook's distance and studentized residuals, the Cook's-distance implementation in the PyOD outlier-detection toolbox [37], empirical cumulative distribution-based outlier detection (ECOD) [38], and the robust regression method TRIM [20]. Because our coordinated attack applies a bounded edit across every stored data entry, within-table discrimination between edited and unedited rows cannot be measured via standard Receiver Operating Characteristic (ROC) curve. Instead, we compare the row rankings obtained from the poisoned data with those obtained from the unpoisoned baseline table, using the diagnostic settings summarized in Section 4.3.

D3 Analytical Model-Data Consistency: D3 evaluates the alignment between candidate models and reported data before and after parameter refitting. First, we compare the document's stated model with the data table. Because $g(x_i)$ is added to both the reference prediction $f_R(x_i)$ and the original recorded output value $y_i$, all fitting residuals remain invariant. Second, we re-estimate the coefficients using unconstrained ordinary least squares (OLS) over the data samples. With a full-rank matrix, the OLS refit absorbs the change into its coefficients and preserves training residuals [25]. Under these assumptions, the original and poisoned training data cannot be distinguished by their fitting residuals alone.

D4 Expression and Prediction Checks: D4 inspects the selected candidate analytical models. Symbolic regression methods constrain operators and reject numerically invalid expressions [39]. Our offline audit parses the equation's Abstract Syntax Tree (AST) to flag nonfinite predictions ($+\infty$, undefined operations), conditional expressions, excessive constant magnitudes, or extreme extrapolation values. Furthermore, negative predictions for non-negative physical quantities trigger physical plausibility flags. But it is to be noted that, because smooth coefficient modifications rarely alter AST structure, these checks alone are insufficient to guarantee security.

Evaluating actual model accuracy requires comparing model outputs against protected, uncorrupted test datasets isolated from the attacker's reach [20]. Protection requires preserving those data outside the attacker's access; a held-out partition of an editable table is insufficient. Section 5.1.1 compares the same selected analytical models using MSE and $R^2$ against poisoned and unpoisoned data, demonstrating how an attack can achieve strong internal fit consistency while failing against physical ground truth.

## 4. Experimental setup

This section describes the manufacturing benchmark, the application of the proposed attack, and the evaluation protocol used to assess their effects compared to analytical model generation workflows under attack-free conditions.

### *4.1 Dataset and reference equation*

The process studied here is Flow assisted Laser-Induced Plasma Micro-Machining (FLIPMM), illustrated in Figure 3. A laser generates plasma near the workpiece to remove material, while flowing water helps carry away debris and bubbles [40]. The process inputs include water speed WS, pulse energy P, pulse repetition frequency F, scanning speed SS and the output is material removal rate (MRR), the volume removed per unit time.

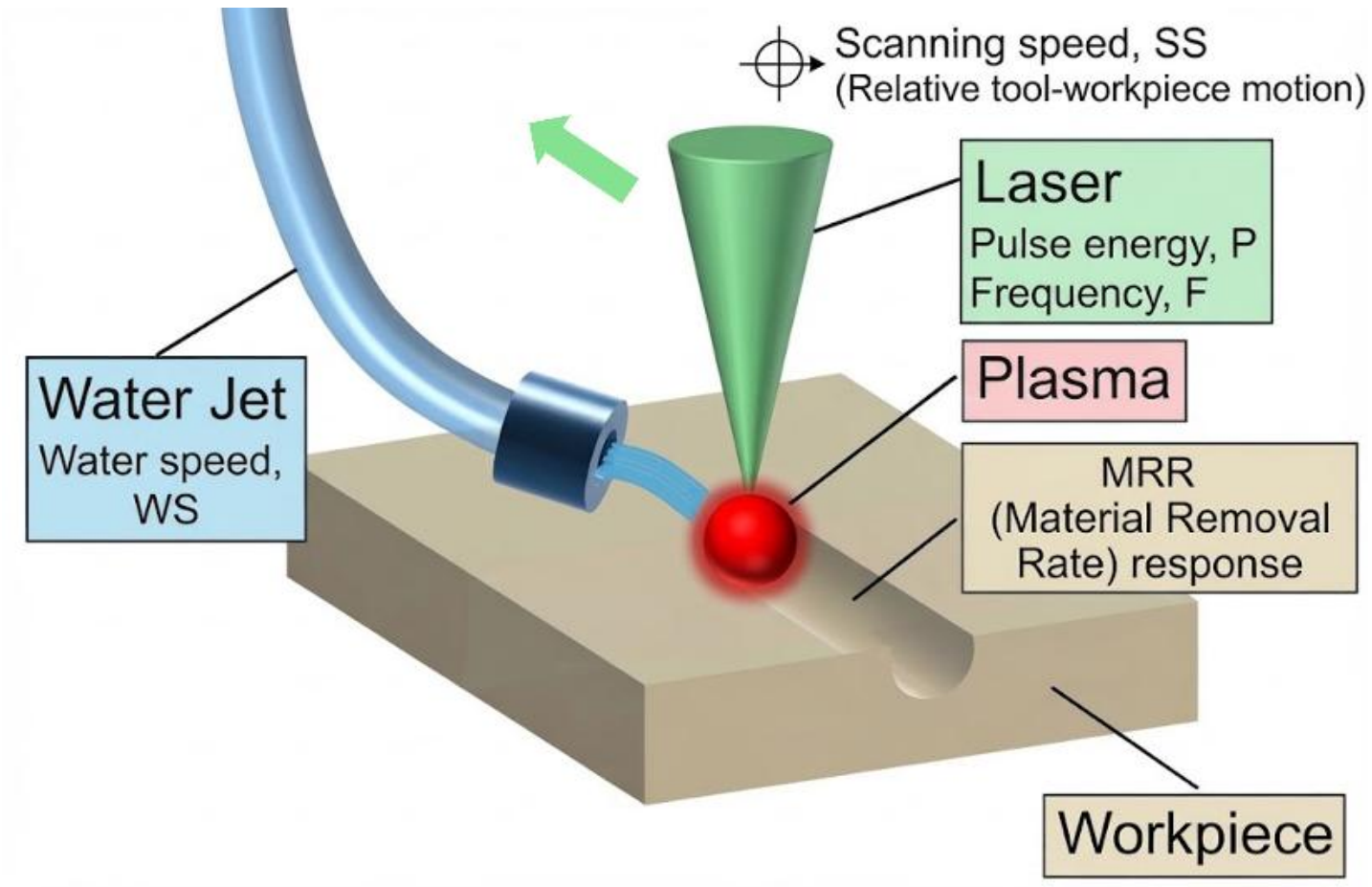

**Figure 3.** Flow assisted Laser-Induced Plasma Micro-Machining (FLIPMM) [40].

The FLIPMM dataset used here comprises 256 settings in a four-level factorial design [15]. The recorded ranges are water speed WS = 6-8 m/s, pulse energy P = 10-30 μJ, pulse repetition frequency F = 2-18 kHz, and scanning speed SS = 1-9 mm/s. Material removal rate is reported in $\mu m^3/s$.

The comparison uses each supplied output directly, with a fixed split of 154 training, 51 validation, and 51 test settings across conditions. The attacker can alter every data output value imported to the workflow, including the training and validation partitions. A clean copy of the test set is saved away from the attacker for evaluating the effectiveness of attacks.

The reference analytical model generated by original framework is fixed before constructing the attacks. We take the five terms from a saved workflow model and fit their coefficients once by ordinary least squares using only the 154 original training data:

$$f_R(W,S,P,FSS) = a_0 + a_1 P + a_2 WSF \\ -a_3/SS + a_4 PF. \qquad (2)$$

*4.2 Attack construction and comparison conditions*

The FLIPMM case applies the coordinated attack to an extrapolation task: predicting MRR beyond the pulse-energy range covered by the dataset. The attacker seeks to make the poisoned model underpredicting MRR by coordinating alterations to the dataset with the poisoned source document.

To apply the coefficient-change construction in Section 3.3 to FLIPMM, we substitute the process variables $x = (W,S,P,FSS)$ and the reference function terms $\phi(x) = (1,P,W,S,F - 1/SSPF)^T$ from Eq. 2 into Eq. 1. With $\Delta\beta = (\Delta,a_0,\ldots\Delta a_4)^T$, the attack on this benchmark is instantiated as:

$$\begin{aligned} & \mid g(x_t) \mid = \mid \Delta a_0 + 40\Delta a_1 + 70\Delta a_2 \\ & - \Delta a_3/5 + 400\Delta a_4 \mid, \\ \text{subject to } & \mid g(x_i) \mid \le \epsilon,\ i = 1,\ldots,256, \\ & g(x_i) = \Delta a_0 + P_i\Delta a_1 + WS_iF_i\Delta a_2 \\ & - \Delta a_3/SS_i + P_iF_i\Delta a_4. \end{aligned} \qquad (3)$$

Since the reference analytical model is linear in P with the other variables fixed, we consider the coefficient vector of the form $\Delta\beta = (\Delta,a_0,\Delta,a_1,000)^T$, altering only the intercept and pulse-energy coefficient. Substituting this vector into $g(x_i) = \phi(x_i)^T\Delta\beta$ gives

$$g(x_i) = \Delta a_0 + P_i\Delta a_1 = g(P), \qquad (4)$$

which produces a straight line in pulse engineering $g(P)$. Solving it with two points $g(10) = +\epsilon$ to $g(30) = -\epsilon$, where the maximum output changes of the data set are supposed to be at the given endpoints of the pulse energy bound, we obtain slope $\Delta a_0 = -\epsilon/10$ and intercept $\Delta a_1 = 2\epsilon$, hence $g(P) = \epsilon(2 - P/10)$.

We evaluate the attack under four prespecified per-entry allowances, $\epsilon$ = 0.5, 1, 2, 4 μm³/s, doubling the permitted alteration between conditions [41]. Figure 4 illustrates this bounded alteration extends beyond the pulse-energy range when $\epsilon$ = 1. The perturbation satisfies $\mid g(P) \mid \le \epsilon$ over 10-30 μJ and at a chosen illustrated sample pulse energy point P = 40μJ, it reaches $g(40) = -2\epsilon$.

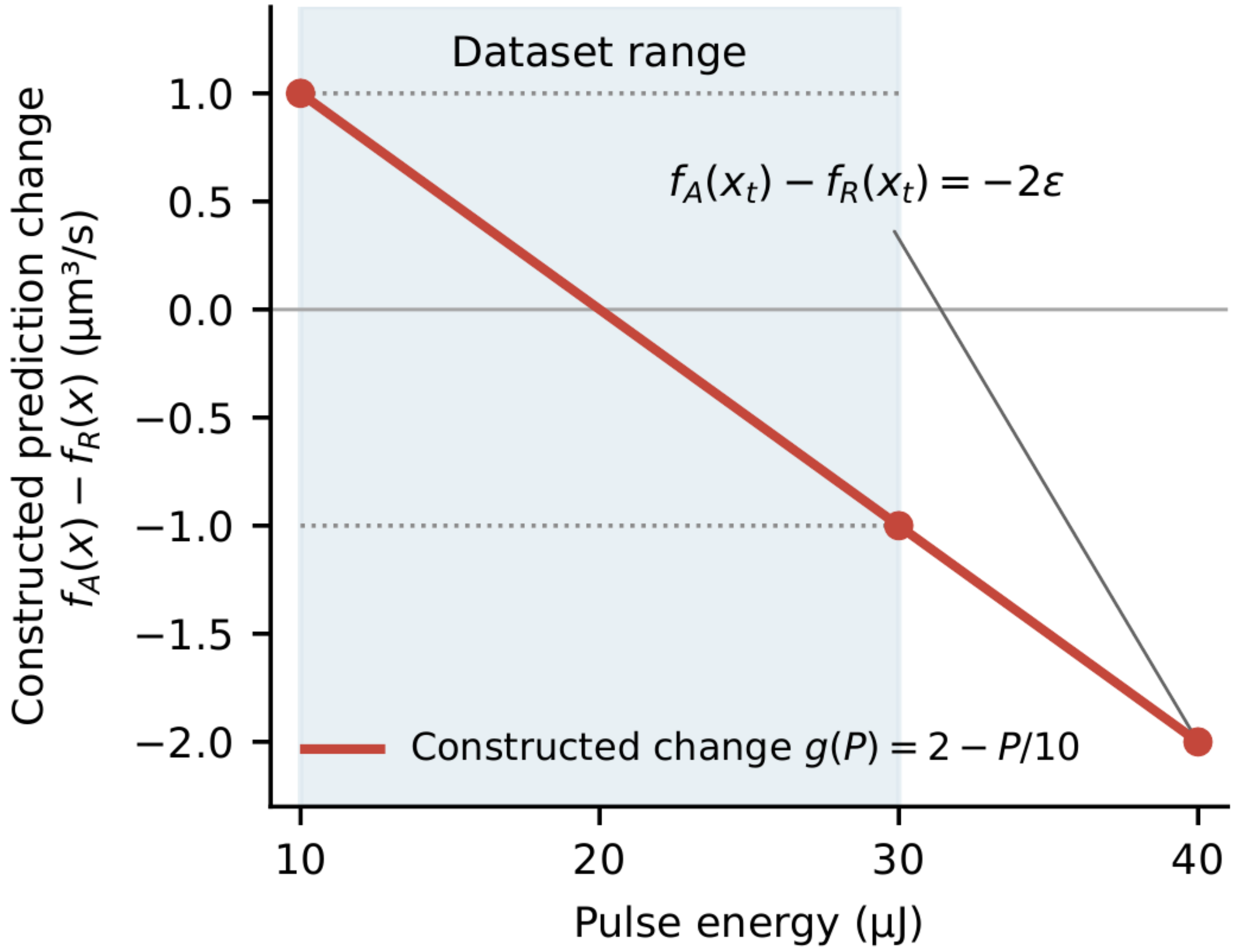

Figure 4. Constructed prediction change before analytical model generation at $\epsilon = 1$ μm³/s. Shading marks the 10-30 μJ calibration range; dotted lines show the edit bound $\pm\epsilon$. At an illustrative extrapolation point P = 40 μJ, the change is $-2\epsilon$.

The document poisoning in Section 3.4 uses the same $\Delta\beta$ to create a poisoned model source document. The document falsely presents $f_A$ as the equation fitted to unmodified training data before the current analysis. For the three model document variants in Section 3.4, I is the source document text without the common heading, and Q is the FLIPMM process query. The query asks how WS, P, F, and SS influence MRR and requests "concrete data or text from the academic papers". Each variant is inserted separately as one additional document alongside the 13 original documents before indexing, so the equation claim is fixed while the inclusion and position of the query change. The full inserted texts are reproduced in A.2.

To separate the effects of the two channels, we compare the original workflow with document-only, data-only, and coordinated poisoning. Document-only poisoning adds the source document while retaining the original numerical data; data-only poisoning edits the MRR values without adding a source document; coordinated poisoning applies both.

*4.3 Evaluation protocol and diagnostic checks*

To evaluate the stealthiness of the attack, diagnostic checks are conducted using D1-D4 from Section 3.6, and measures prediction accuracy against original unpoisoned benchmark data. All model-generation conditions use GPT-4o mini for synthesis and equation proposals, text-embedding-3-small for retrieval, and BAAI/bge-reranker-large for reranking. The analytical model searches allow 51 initial proposals of analytical models and followed by up to ten refinement rounds with 21 additional proposals each, with a validation-$R^2$ stopping threshold of 0.98.

For the document channel, we assess both propagation of the altered claim and document plausibility. We record the poisoned document's inclusion and rank among the five context passages and whether the knowledge summary exactly repeats its printed pulse-energy coefficient and predicted MRR at $x_t$. Document Plausibility (D1) is assessed separately by GPT-4o mini at temperature 0, which judges each poisoned document alongside the 13 original documents without receiving $f_R$. We report the mean and range of 0-1 suspicion scores across eight matched document orders, with larger values indicating greater suspicion. The three source document variants at $\epsilon$ = 1 (24 assessments) and documentt alone across four bounds (32 assessments) are summarized separately.

For the numerical channel, at each bound, the six D2 Response Anomalies rank the 205 training and validation rows, using the functions in Eq. 2 for regression and excluding test data. Spearman correlation with the original-table rankings measures changes in review priority; 1 means unchanged rankings. D3 Analytical Model-Data Consistency compares $f_R$ with original MRR values and $f_A$ with poisoned values through mean absolute error (MAE) over all 256 settings and changes in individual residuals. The comparison is repeated after unconstrained OLS refitting of the five fixed functions to each table's 154 training rows.

Because model generation may select an equation different from $f_A$, D4 Expression and Prediction Checks examines the initial and final selected equations. Structural checks flag conditional or piecewise expressions, comparisons, and constants with magnitude at least 1,000. Prediction checks flag nonfinite or negative MRR and magnitudes above ten times the largest absolute original MRR value. The prediction checks use a 256-setting grid with four equally spaced values per variable over WS = 6-9 m/s, P = 10-35 μJ, F = 2-22 kHz, and SS = 1-9 mm/s.

To assess prediction accuracy independently of the poisoned data, we evaluate the same selected equation against between poisoned and original MRR values without refitting. MSE and $R^2$ are calculated per partition and averaged across successful searches within each condition. Lower MSE indicates closer predictions; $R^2$ = 1 is perfect prediction and 0 matches the partition's mean-MRR baseline.

## 5. Results

This section first presents the effects of poisoning on selected equations through overall prediction results, channel ablation, and the effect of the edit bound (Section 5.1). It then examines detectability under D1-D4, relating each diagnostic result to the information available to the check (Section 5.2).

### *5.1 Effects of poisoning on analytical model generation*

We compare the analytical models selected under clean, document-only, data-only, and coordinated conditions to assess how poisoning affects analytical model generation.

#### *5.1.1 Overall prediction results*

The coordinated-poisoning results show a similar trend in the final prediction accuracy across all four edit bounds $\epsilon$. We use $\epsilon = 4$ as an illustrative example in Table 2, which compares the initial and final selections for the original workflow and all three attack conditions. Complete results for all $\epsilon$ studies are provided in A.1. The analytical models selected under coordinated poisoning are less accurate on the unpoisoned test dataset after refinement than those from the original workflow. In Table 2, final clean-test MSE is 71.37 under coordinated poisoning and 50.27 under the original workflow, with corresponding $R^2$ values of 0.789 and 0.851.

Table 2. Mean $R^2$ and MSE before (Initial) and after (Final) refinement at $\epsilon = 4$ μm³/s, alongside the original clean baseline. Poisoned validation reports results on the attacker's altered validation data; clean validation evaluate reports results on the original unpoisoned validation data. Test evaluate results on the unpoisoned test set away from the attacker access. Dashes indicate conditions without data poisoning.

| Condition | Stage | Poisoned validation | | Clean validation | | Test | |
|---|---|---|---|---|---|---|---|
| | | $R^2$ | MSE | $R^2$ | MSE | $R^2$ | MSE |
| Clean | Initial | – | – | -0.128 | 403.88 | -0.067 | 360.86 |
| Clean | Final | – | – | 0.871 | 46.06 | 0.851 | 50.27 |
| Document only | Initial | – | – | 0.834 | 59.29 | 0.796 | 68.92 |
| Document only | Final | – | – | 0.870 | 46.58 | 0.825 | 59.06 |
| Data only | Initial | -0.065 | 335.73 | 0.040 | 344.01 | 0.103 | 303.30 |
| Data only | Final | 0.872 | 40.37 | 0.855 | 52.11 | 0.821 | 60.46 |

| Condition | Stage | Poisoned validation | | Clean validation | | Test | |
|---|---|---|---|---|---|---|---|
| | | $R^2$ | MSE | $R^2$ | MSE | $R^2$ | MSE |
| Coordinated | Initial | 0.828 | 54.34 | 0.794 | 73.64 | 0.758 | 81.74 |
| Coordinated | Final | 0.862 | 43.43 | 0.824 | 63.05 | 0.789 | 71.37 |

The same selected analytical models receive more favorable validation scores against poisoned data. Under coordinated poisoning, final poisoned-validation MSE is 43.43 and $R^2$ is 0.862, compared with clean-validation MSE of 63.05 and $R^2$ of 0.824 for the same analytical models, predictions, and 51 validation inputs. Clean validation has higher MSE and lower $R^2$ in all 15 coordinated runs. Validation feedback based on poisoned output values therefore conceals part of the prediction error relative to benchmark ground-truth values.

Figure 5 extends the comparison across all four edit bounds, showing mean paired differences between each attack condition and the original workflow after refinement. Panel (a) reports changes in clean-test MSE to assess prediction accuracy on unpoisoned benchmark data. Panel (b) reports changes in predicted MRR at a point P = 40 μJ, the illustrative extrapolation case beyond the dataset's pulse-energy range.

At $\epsilon = 4$, 14 pairs of runs from the original workflow and coordinated poisoning give a mean clean-test MSE increase of 20.90 $(\mu m^3/s)^2$, with an exploratory 95% interval of [9.56, 32.24] (Figure 5(a)). These paired differences use only repetitions with valid models in both conditions, while Table 2 averages all successful runs within each condition. With the fixed clean-test output values, this increase corresponds to a mean $R^2$ decrease of 0.062. Preserved consistency between the poisoned analytical model and poisoned numerical data (Section 5.2.3) therefore cannot rule out poorer predictions than those from the original workflow.

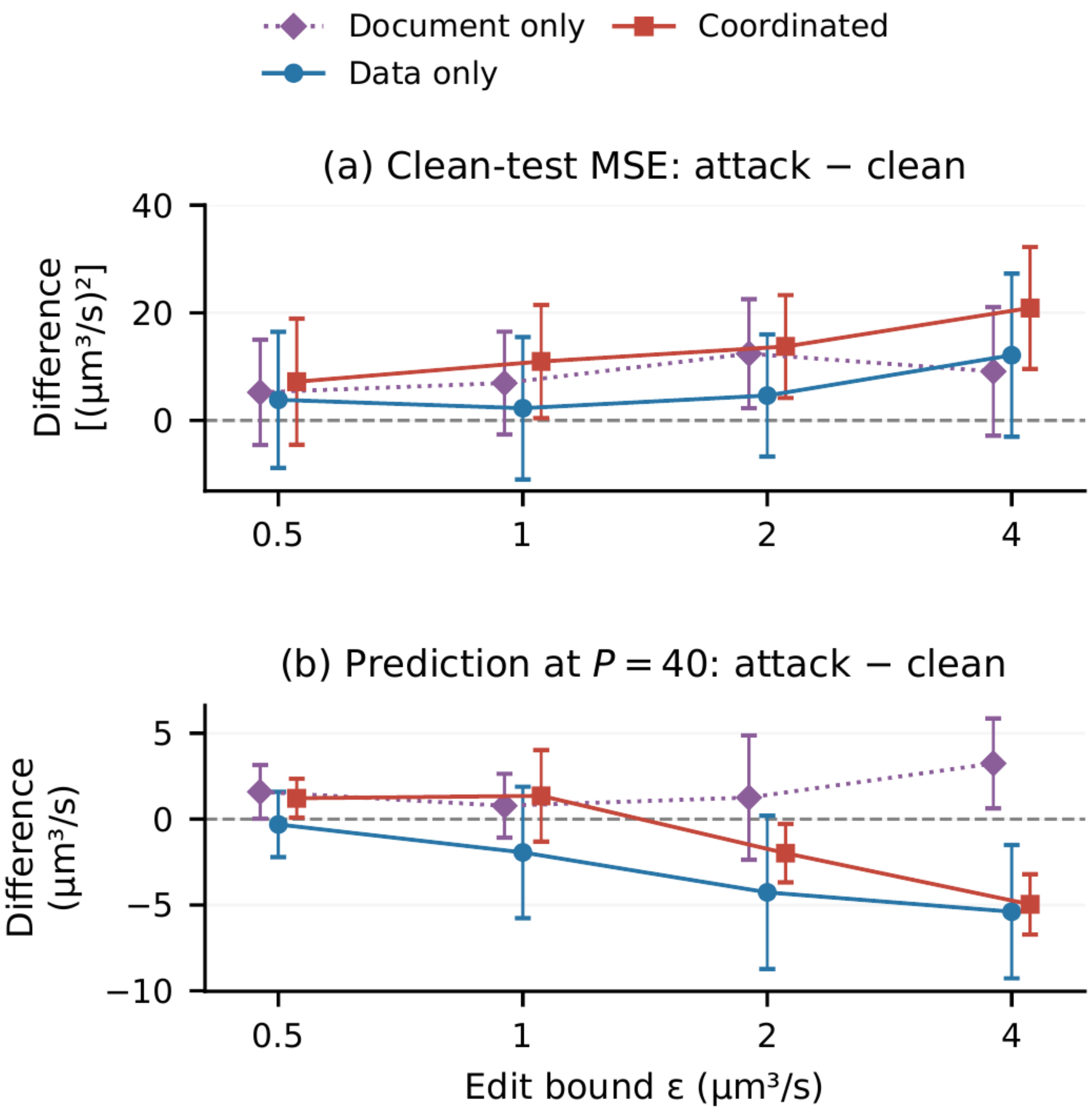

Figure 5. Effects of targeted poisoning after model selection: (a) clean-test MSE. Points show mean paired differences (attack results minus original workflow results); bars show exploratory, unadjusted 95% Student-t confidence intervals. Runs from the original workflow are reused across bounds, which double between ticks and (b) predicted MRR in the illustrative extrapolation case at P = 40 μJ.

### *5.1.2 Ablation of the document and data channels*

To examine the contribution of each channel (document or data alteration), we compare coordinated poisoning with document-only and data-only cases. Coordinated poisoning produces higher clean-test error than data-only poisoning at $\epsilon = 4$ (Figure 5(a)). Holding the numerical alterations fixed, adding the poisoned source document gives a mean paired MSE increase of 10.74 $(\mu m^3/s)^2$ over 14 pairs of runs, with a 95% interval of [1.76, 19.72] and a corresponding mean $R^2$ decrease of 0.032. The intervals at the lower edit bounds include zero, limiting the evidence for this additional clean-test loss to the largest tested allowance.

The single-channel comparisons with the original workflow are less consistent. Document-only poisoning increases mean clean-test MSE by 12.41 $(\mu m^3/s)^2$ at $\epsilon = 2$, with interval [2.27, 22.54]; its intervals at the other bounds include zero. The data-only intervals include zero at every bound (Figure 5(a)). At the illustrative extrapolation input P = 40 μJ, the comparisons across the four edit bounds do not establish a greater change in predicted MRR under coordinated poisoning than under data-only poisoning.

Document-only poisoning also does not establish a decrease relative to the original workflow at any bound (Figure 5(b)). We also compare the prediction change at P = 40 μJ under coordinated poisoning with the sum of the single-channel changes. The intervals at all four edit bounds include zero, so these comparisons do not establish an additional downward effect.

The retrieval check confirmed that the poisoned source document entered the LLM's context in all 120 primary runs containing poisoned source documents; supporting retrieval and claim-repetition results are reported in A.2. The document comparisons measure the overall effect of inserting a poisoned source document. They do not isolate the effect of altering its numerical claim from that of adding a source document.

### *5.1.3 Effect of the edit bound*

The observed mean clean-test loss under coordinated poisoning increases with the edit bound (Figure 5(a)). The intervals at $\epsilon = 1, 2, 4$ lies above zero, whereas the interval at $\epsilon = 0.5$ includes zero. The same runs of the original workflow are reused at every bound; these comparisons do not test differences between the edit bounds.

Coordinated poisoning decreases the final prediction at P = 40 μJ at the two larger bounds (Figure 5(b)). At $\epsilon = 4$, its mean difference from the original workflow is −4.96 μm³/s with [−6.72, −3.20]. Downward movement occurs in 12 of 14 pairs of runs at both $\epsilon = 2$ and 4. The smaller bounds do not establish a downward mean change. These are effects after fitting and selection; they need not equal the constructed decrease of $2\epsilon$ relative to the saved reference.

The clean-test accuracy loss accompanies an improvement at P = 40 μJ. All predictions from coordinated runs at P = 40 μJ remain above the benchmark polynomial prediction of 39.813, so lowering them brings them closer to that reference. Following the attack objective therefore does not imply increased error at the extrapolation point.

## *5.2 Detectability*

We next interpret the D1-D4 findings under the information-access assumptions in Section 3.6. The subsections below explain what document ratings, row rankings, internal fit, and the validity of selected analytical models reveal about the attack. Suspicion scores and ranking correlations remain descriptive; "pass" is used only for specified binary checks.

#### *5.2.1 D1: Document Plausibility*

At $\epsilon = 1$, the question-document and document-question variants receive higher mean suspicion scores than document alone, even though their analytical model and predicted MRR claim are identical (Table 3). The low-scoring assessment describes a plausible but unvalidated analytical model, whereas the high-scoring assessment cites the heading stating that the document is synthetic. These explanations concern plausibility and provenance of the document variants; rather than focusing on the recognition of the coefficient substitution.

A separate assessment of document alone gives mean suspicion scores of 0.338 − 0.338 across the four edit bounds from eight orders per bound. Increasing the permitted numerical alteration therefore does not increase mean suspicion in this cohort.

Table 3. D1: mean and minimum-maximum range of suspicion scores across eight document orders per model source document variant at $\epsilon = 1$ μm³/s. The 0-1 scale indicates increasing suspicion; no rejection threshold is defined.

| Presentation | Mean suspicion score | Observed range |
|---|---|---|
| Document alone | 0.388 | 0.1–0.9 |
| Question-document | 0.613 | 0.2–1.0 |
| Document-question | 0.525 | 0.3–0.9 |

#### *5.2.2 D2: Data Anomalies*

The original and poisoned row rankings remain strongly correlated across all four edit bounds (Table 4). Cook's distance, studentized residuals, and TRIM retain identical rankings; correlations for Cook's distance in PyOD and for ECOD exceed 0.995, while cleanlab spans 0.918-0.998.

The poisoning changes every output value while largely preserving review priorities. However, because a table-rejection rule is not considered within this experiment context, the acceptance of the altered data is not established from these correlations.

Table 4. D2: Spearman correlations between original and poisoned row rankings for 205 training/validation rows. Ranges span $\epsilon = 0.5, 1, 2, 4$ μm³/s; 1 means unchanged review priority. Correlations are not detection rates.

| Diagnostic score | Spearman rank correlation |
|---|---|
| cleanlab | 0.918–0.998 |
| Cook's distance (OLS) | 1.000 |
| Studentized residuals | 1.000 |
| Cook's distance (PyOD) | 0.997–1.000 |
| ECOD | 0.996–1.000 |
| TRIM | 1.000 |

#### *5.2.3 D3: Analytical Model-Data consistency before and after refitting*

The MAE, evaluated over all 256 data samples, remains 5.305 µm³/s at every edit bound, $\epsilon$ = 0.5, 1, 2, 4: the reference analytical model agrees with the original table to the same extent that the altered analytical model agrees with the poisoned table. Refitting the coefficients of the five fixed reference functions by unconstrained OLS leaves this MAE unchanged. Across both checks, the largest change in any individual residual is below $10^{-12}$ µm³/s, consistent with numerical roundoff.

The attack therefore preserves the reference family's existing fitting error. Adding the same $g(x_i)$ to the predicted MRR and MRR values in the dataset leaves no additional residual signal, including after refitting. A fixed MAE threshold that accepts the original analytical model and table consequently also accepts the altered pair, apart from roundoff. The unchanged MAE measures agreement with the corresponding data table; it does not independently validate the poisoned MRR values. This finding concerns the fixed reference functions before analytical model generation.

#### *5.2.4 D4: Expression and Prediction Checks*

The selected analytical models may reveal problems beyond the preserved fit of the reference family. Across the 386 initial and final evaluations from 193 successful runs, none triggers the specified structural flags or produces nonfinite or extreme predictions on the diagnostic grid (Table 5).

The nonnegativity result is different: in 290 evaluations, the selected analytical model predicts negative material removal rate at one or more grid settings. Final-stage failures include all 60 document-only and all 60 coordinated selections, alongside 6 of 14 clean selections (Table 5). Because analytical models from original workflow runs also fail on this extended grid, the warning identifies a limitation of the selected analytical models without establishing poisoning as its cause.

Table 5. D4: evaluations with a negative grid prediction before (Initial) and after (Final) refinement, shown as failures/total. Attack rows pool four bounds. All 386 evaluations pass structural checks and verify for nonfinite or extreme predictions.

| Condition | Initial | Final |
|---|---|---|
| Clean | 5/14 | 6/14 |
| Document only | 58/60 | 60/60 |
| Data only | 17/59 | 26/59 |
| Combined | 58/60 | 60/60 |

## 6 Conclusion

This work demonstrated that adversarial attacks could compromise automated modeling pipelines by simultaneously altering retrieved text claims and experimental data. The proposed dual-channel coordinated attack ensures that the alteration in the generated analytical model survives standard numerical fitting and validation checks. We demonstrate the proposed adversarial attack in an LLM-driven, retrieval-assisted manufacturing predictive modeling framework by constructing a poisoned document and bounded edits on the training dataset. We then evaluate attacks' impact by comparing the analytical models generated by the poisoned framework against clean benchmark data. The key findings are summarized as following:

1) Data-only poisoning worse the final chosen output prediction at the largest edit bound, showing that bounded data edits can influence analytical models after fitting and refinement.
2) Document-only poisoning propagated altered numerical claims into the modeling context but did not establish a decrease in the final prediction at the process-variable values chosen by the attacker.
3) Coordinated poisoning preserved residuals within the reference analytical model family while degrading clean-test accuracy after model generation. At the largest edit bound, coordinated poisoning produced greater clean-test error than data-only poisoning.
4) Through diagnostic evaluation, document plausibility varied with presentation and document order, while data-anomaly rankings changed little. Coordinated alterations preserved residuals even after coefficient refitting, showing that internal diagnostics could not independently verify integrity. Negative-prediction warnings also occurred in original clean runs, so those warnings alone did not identify poisoning.

Though the proposed adversarial attack approach is generally applicable to various LLM-driven frameworks, this study is limited to one numerical benchmark, a specific RAG-LLM modeling framework, and a threat model permitting full data-table modifications. Future work will extend this evaluation to broader manufacturing processes, model generators, and partial data-poisoning scenarios. We will also develop defense mechanisms for LLM-driven process model generation based on the proposed attacks.

**Acknowledgments**

The authors gratefully acknowledge financial support from the National Science Foundation (CMMI-2434384, CNS-2308730, and CMMI-2326341). HAN gratefully acknowledges the Pratt & Whitney Institute for Advanced Systems Engineering Fellowship from the University of Connecticut. HX also acknowledges the support of the Pratt & Whitney Associate Professorship in Advanced Systems Engineering of the University of Connecticut. Ziqin Liu was supported by National Science Foundation Research Traineeship (NRT) grant #2152202.

## Appendix A. Supplementary materials

### *A.1 Supplementary prediction results*

Supplementary tables reporting mean $R^2$ and MSE before and after refinement at $\epsilon$ = 0.5, 1, 2, 4 μm³/s are available at https://github.com/hoanganh-2704/rag-analytical-model-poisoning-supplement/blob/main/epsilon_results.md.

### *A.2* ***Document retrieval and numerical claim repetition***

The verbatim poisoned source document texts are available in the accompanying repository documentation at https://github.com/hoanganh-2704/rag-analytical-model-poisoning-supplement/blob/main/poisoned_document_texts.md. The document contains the document-alone, question-document, and document-question variants, preserving their original wording and numerical precision. Table A.1 summarizes their retrieval and numerical claim repetition.

Table A.1. Retrieval and numerical claim repetition at $\epsilon$ = 1 μm³/s (15 repetitions per condition). The three variants share the same poisoned model text. Retrieval counts inclusion among five context passages; repetition counts exact matches to the printed values. A dash indicates no added text.

| Presentation | Retrieved into context | Value repeated in summary | |
|---|---|---|---|
| | | Pulse-energy coefficient | Prediction at $P$ = 40 μJ |
| Document alone | 15/15 | 15/15 | 0/15 |
| Question-document | 15/15 | 15/15 | 10/15 |
| Document-question | 15/15 | 15/15 | 0/15 |
| Original corpus | — | 0/15 | 0/15 |

In the primary comparison, all 120 runs with poisoned documents retrieve the document and repeat its pulse-energy coefficient. Repetition of the predicted MRR at $x_t$ varies with question placement. These counts concern exact repetition of eight-decimal values in answers to one fixed process question, with one fixed index per condition. A repeated value does not establish endorsement of the claim, general attack success, or adoption by the selected equation. These propagation results are distinct from the selected-model results in Section 5.1, which use document alone throughout.